\documentclass[twocolumn]{aastex631} 

\usepackage{graphics,epsf}
\usepackage[utf8]{inputenc}
\usepackage{amsmath}                
\usepackage{amsfonts}               
\usepackage{amssymb}                
\usepackage{epsfig}                 
\usepackage{graphicx}               
\usepackage{float}
\usepackage{color}
\usepackage{multirow}               
\usepackage{hyperref}

\hypersetup{
    colorlinks=true,
    linkcolor=red,   
    urlcolor=cyan}

\usepackage[colorinlistoftodos]{todonotes}

\newcommand{\s}{{~\rm s}}

\newcommand{\g}{{~\rm g}}
\newcommand{\G}{{~\rm G}}
\newcommand{\K}{{~\rm K}}
\newcommand{\erg}{{~\rm erg}}

\begin{document}

\title{Et tu, Brute?: The Crab Nebula also exploded by jittering jets}
\date{July 2024}

\author[0000-0002-9444-9460]{Dmitry Shishkin}
\affiliation{Department of Physics, Technion, Haifa, 3200003, Israel; s.dmitry@campus.technion.ac.il; soker@physics.technion.ac.il}

\author[0000-0003-0375-8987]{Noam Soker}
\affiliation{Department of Physics, Technion, Haifa, 3200003, Israel; s.dmitry@campus.technion.ac.il; soker@physics.technion.ac.il}

\begin{abstract}

We identify a point-symmetrical morphology comprised of seven pairs of opposite bays in the core-collapse supernova (CCSN) remnant Crab Nebula, which is consistent with the jittering jets explosion mechanism (JJEM) of CCSNe.
We use a recently published infrared image of the Crab Nebula and apply image analysis to fit seven pairs of bays in the Crab, each pair of two bays and a symmetry axis connecting them. The seven symmetry axes intersect close to the explosion site, forming a point-symmetrical structure. We explain the bays as clumps that move slower than the low-density ejecta that the pulsar accelerated. Jittering jets that exploded the Crab formed the clumps during the explosion process. This shows that jittering jets explode even very low-energy CCSNe, as the Crab is, adding to the solidification of the JJEM as the primary explosion mechanism of CCSNe. 
\end{abstract}

\section{Introduction}
\label{sec:Introduction}

The identification of point-symmetrical morphology in eleven core-collapse supernova (CCSN) remnants (CCSNRs; e.g., references in \citealt{Bearetal2025, Soker2025W44}) strongly supports the jittering jets explosion mechanism (JJEM; e.g., \citealt{Soker2024CF, Soker2024Keyhole, Soker2025Learning, WangShishkinSoker2024} for recent studies of the JJEM) as the primary explosion mechanism of CCSNe. The competing neutrino-driven explosion mechanism (e.g., \citealt{Andresenetal2024, Burrowsetal2024kick, JankaKresse2024, Mulleretal2024, Nakamuraetal2024} for some recent studies), cannot account for these point-symmetrical morphologies \citep{Soker2024Rev, SokerShishkin2024Vela}. This motivates the search for point-symmetrical morphologies in more CCSNRs. We examine the Crab Nebula.   

Many studies over the years explored the structure of the Crab Nebula, observationally (e.g., \citealt{Clarketal1983, Fesen_etal_1992, BietenholzNugent2015, Dubner_VLA_2017, Martin_3D_2021, Temimetal2024}) and theoretically (e.g., \citealt{Porthetal2014, BlondingChevalier2017, DirsonHorns2023}). The explosion energy of the Crab Nebula is very low, $E_{\rm exp} \simeq 5 \times 10^{49} -10^{50} \erg$ (e.g., \citealt{YangChevalier2015}) although \cite{BietenholzNugent2015} argue for a somewhat larger explosion energy.   
At the same time, the energy that the pulsar deposited into the pulsar wind nebula (PWN) is $E_{\rm P} \simeq 3.5 \times 10^{49} \erg$ (e.g., \citealt{YangChevalier2015}). The relevant point to our study is that $E_{\rm P}$ is non-negligible relative to $E_{\rm exp}$, implying that the PWN has substantially influenced the expanding ejecta (e.g., \citealt{BietenholzNugent2015, BlondingChevalier2017}). 

In a recent study, \cite{Temimetal2024} marked nine indentations in the Crab Nebula similar to the larger bays in the east and west (e.g., \citealt{Michaeletal1991, Dubner_VLA_2017}). 
\cite{Michaeletal1991} identified only the two opposite east and west bays and suggested that they are the projection of a torus on the plane of the sky. However, with the identification of seven more bays by \cite{Temimetal2024} and five in this study, this explanation can only hold for some bays or none. Instead, we note that six of the nine bays that \cite{Temimetal2024} mark form three pairs and that their three symmetry axes (the lines connecting two opposite bays) cross at the same point, coinciding with the projected location of the pulsar at the explosion. In Section \ref{sec:BaysObs}, we identify more pairs, and in Section \ref{sec:Bays}, we establish the point-symmetrical morphology of the Crab Nebula. In Section \ref{sec:Summary} we suggest that the bays are dense clumps that the PWN flows around, and we summarize this \textit{Letter} by further strengthening the JJEM as the primary explosion mechanism of CCSNe.

\section{Identifying pairs of bays in recent observations}
\label{sec:BaysObs}

\cite{Temimetal2024} perform a detailed analysis of the Crab Nebula based on JWST \citep{Gardner_JWST} observations, including several filtered images taken using the MIRI \citep{Wright_MIRI} and NIRCam \citep{Reike_NIRCam} instruments.
In these images, for a wide range of filters, including F162M, F480M (NIRCAM) and, F560W, F1130W (MIRI), indentations in the SNR, which are termed bays, can be easily identified; most prominent in the F480M image ($460-500 \rm nm$) that shows the synchrotron emission \citep{Temimetal2024}. 
We chose this image as the basis for our study.

\begin{figure*}
\begin{center}
\includegraphics[trim=1cm 0cm 1cm 0cm,width=\textwidth]{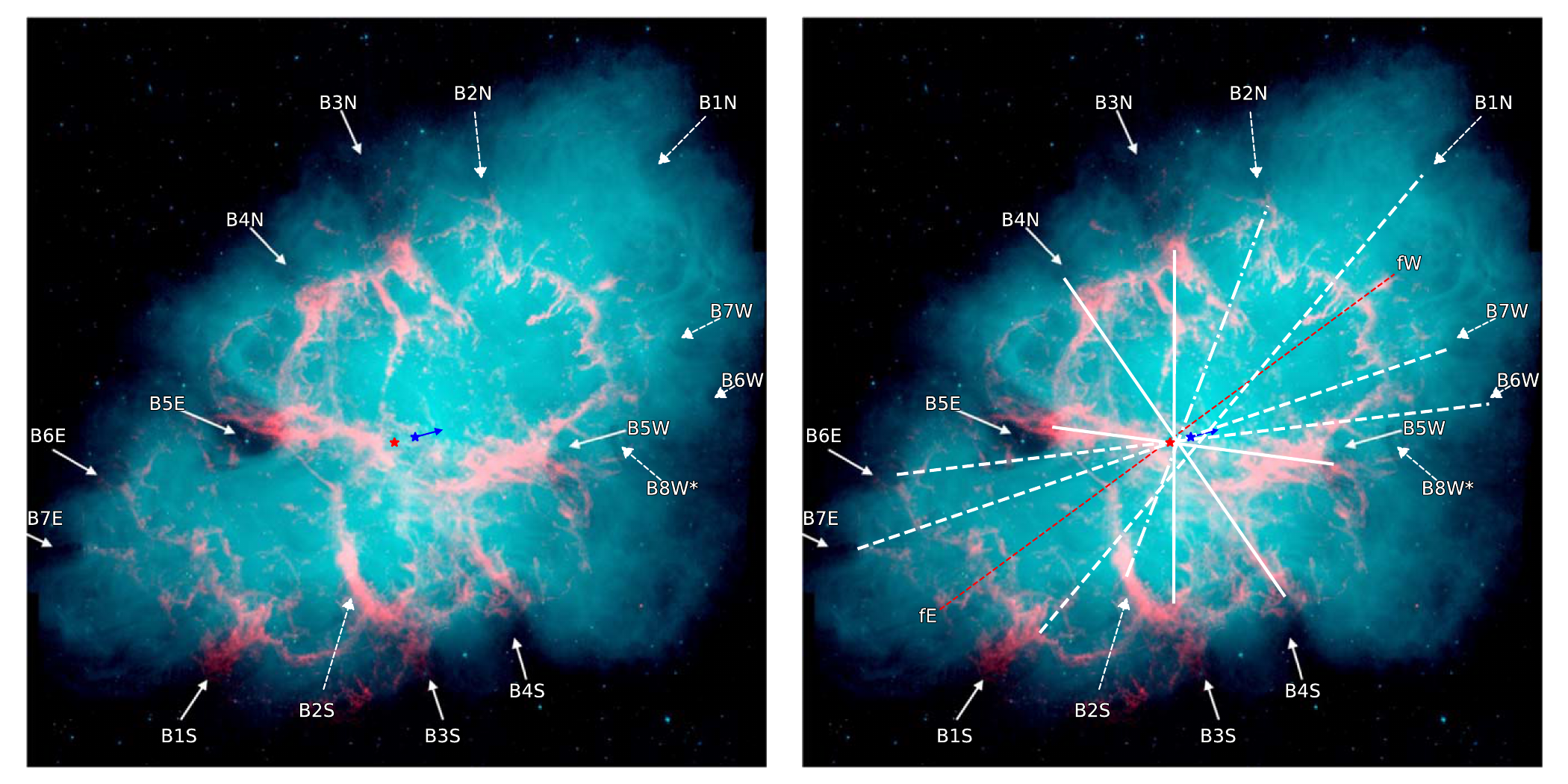} 
\caption{An image of the Crab Nebula adapted from Figure 13 of \cite{Temimetal2024}. Blue is synchrotron, and red is dust emission. \cite{Temimetal2024} identify nine bays with dust filaments that they mark with solid-white arrows. We mark five additional bays that we identify in this image with dashed-white arrows. Three solid lines connect bay pairs from bays identified by \cite{Temimetal2024}, while four dashed lines connect bay pairs that include new bays identified by us. A red dashed line depicts a possible filament pair. 
We denote with a blue asterisk the present location of the Crab Pulsar (PSR B0531+21), with a red asterisk the calculated location of the neutron star (NS) at the explosion, and with a blue arrow the NS direction of motion based on \cite{Kaplan_etal_2008}. We note that other, more current estimates give a lower proper motion estimate, especially in the declination \citep{GAIA_DR3_2023, Lin_VLBI_2023}. 
The edges of the symmetry axes of the bays (the line connecting bays) in the right panel are at the focal points of the fitted parabolas to the bays that we calculate in Section \ref{sec:Bays}.}
\label{fig:Temim2024}
\end{center}
\end{figure*}

In Figure~\ref{fig:Temim2024}, we present an annotated version of Figure 13 from \cite{Temimetal2024}, where blue indicates the synchrotron and red dust emissions. In this figure, they mark nine bays (indentations in the synchrotron emission, blue) that coincide with dust filaments (red) with solid arrows.
We noticed six of the nine bays form pairs forming a point-symmetrical structure. We connect these bays by three solid-white lines in the right panel of Figure~\ref{fig:Temim2024}.  
We identify five more bays that we point at with dashed arrows; one bay coincides with a dust filament (B2N). 
We connect these additional bays with dashed-white lines. 
We also identify two large dust filaments (fE and fW) that we connect with a dashed-red line in the right panel of Figure~\ref{fig:Temim2024}. 
The original three pairs of the bays that \cite{Temimetal2024} identified (solid-white lines), the four pairs with the bays we identify we denote here (dashed-white lines), and the pair of filaments (dashed-red line) form a point-symmetrical structure (right panel of Figure~\ref{fig:Temim2024}). 

Previous studies have already identified the pair of bays B5, which form the closest pair and largest two bays (B5E and B5W in Figure~\ref{fig:Temim2024}), in the optical and infrared (IR; e.g., \citealt{Fesen_etal_1992}).
\cite{Fesen_etal_1992} also identified an additional bay that we mark B8W* in Figure~\ref{fig:Temim2024}, but found it to have some different properties from the largest bays. We mark the bay in our figures but leave its analysis to future studies. We suspect that the opposite bay of B8W* is very close to bay B5E, southeast of it. Therefore, to separate this possible opposite bay requires a deeper analysis that we leave for a future study. 

The largest bays are also visible in X-ray emission, e.g., \cite{Seward_Chandra_2006}, who denote the two bays as East and West. They also mention numerous complicated X-ray structures at the edge of the nebula. \cite{Dubner_VLA_2017} notice that this pair of bays is less prominent in radio, relative to IR observations (Spitzer, at $8 \rm\mu m$ and $4.5 \rm \mu m$, \citealt{Temim_Spitzer_2006}) and optical (HST F547M, at $510-590 \rm nm$, \citealt{Loll_HST_2013}).  \cite{Dubner_VLA_2017} highlight the presence of radio filaments in the two largest bays (B5) and in the ``arcade of loops'' region, B4S in Figure \ref{fig:Temim2024}. 

\cite{Martin_3D_2021} reconstructed the three-dimensional structure of the Crab Nebula in line emission flux. In the $H\alpha$ emission map, they revealed two ``depressions'' coinciding with the largest bays (B5E and B5W) and an indentation in the vicinity of the ``arcade'' that \cite{Dubner_VLA_2017} identified (B4S in Figure~\ref{fig:Temim2024}). 

Our main result is that the largest bays form one pair out of seven pairs of bays that form the point-symmetrical morphology. 

\section{The point-symmetrical bays}
\label{sec:Bays}

We use JWST NIRCam F480 data (Program ID: 1714; PI: Dr. Tea Temim), retrieved from the MAST archive.
We use the level 3 image product (hereafter ``JWST image'') as provided and do not pass the image through any additional processing pipelines besides those described below.

\begin{figure}[t]
\begin{center}
\includegraphics[trim=2cm 0cm 1cm 1cm, width=0.5\textwidth]{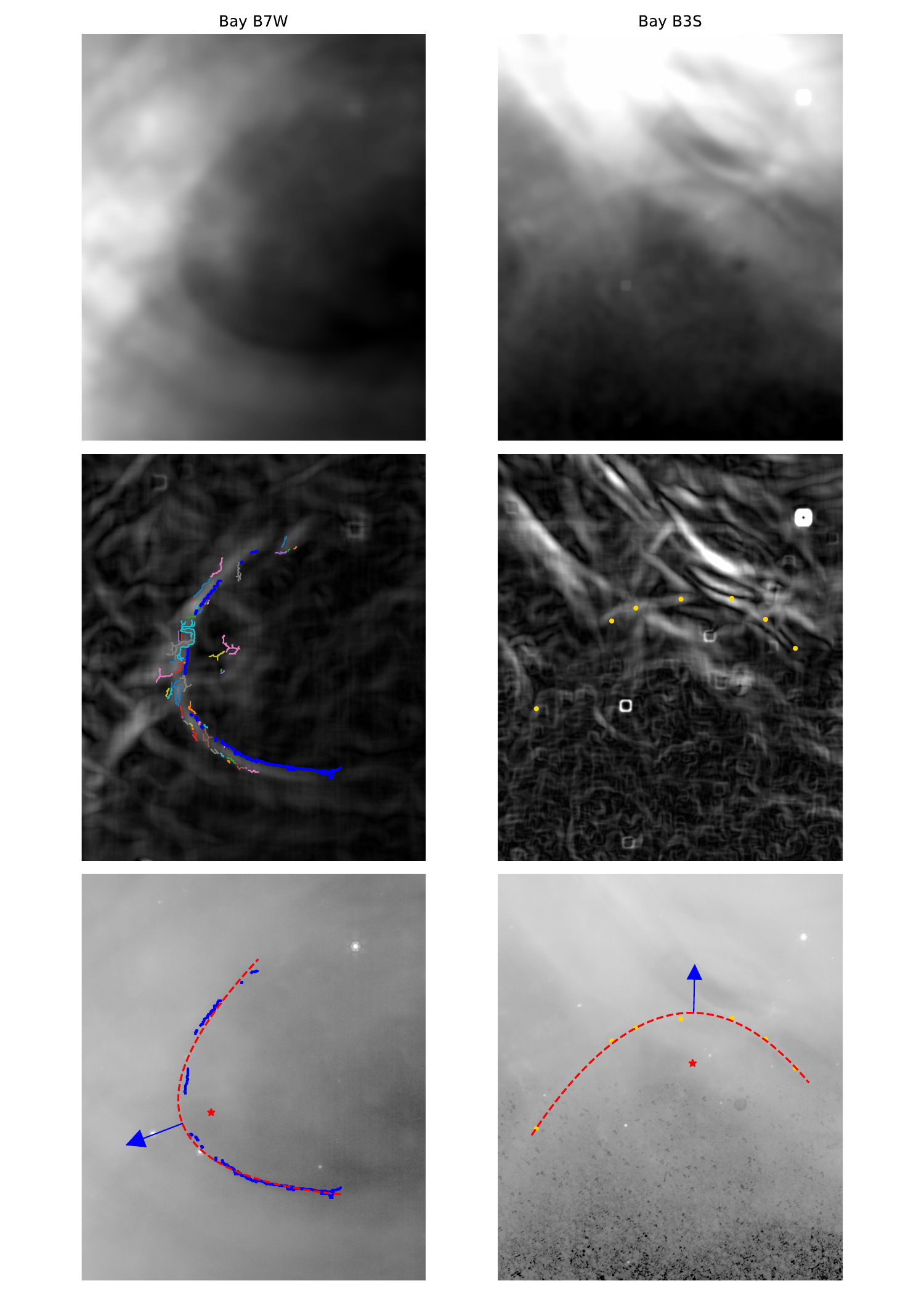} 
\caption{Description of analysis pipeline for bay definition and fitting for two selected bays: B7W (left column) and B3S (right column). \textbf{Top panels:} source removed, Gaussian and median blurred version of the JWST image (smoothed, source-removed), centered around the bays. \textbf{Middle panels:} Sobel-processed image of the smoothed source-removed image, emphasizing gradients. We also denote identified edges with multi-colored lines and those selected for fitting purposes with blue lines. Gold dots indicate manually selected bay-defining points where automatic edge detection yielded poor results (right-middle panel). \textbf{Bottom panels:} JWST image, where we also denote the edges/points used for the parabolae fitting (in blue/gold), the fitted parabola (dashed red), the parabola focal points (red asterisk) and parabola pointing (blue arrow).}
\label{fig:BayDetails}
\end{center}
\end{figure}

To define individual bays, we first define a polygon for each bay we suspect. In each polygon, point sources that might interfere with the definition are manually identified and removed with designated Python code. 
We use the OpenCV library \citep{opencv_library} to process each bay region individually. The following methods and algorithms are from the OpenCV library. We pass the image through Gaussian and median smoothing to reduce noise, as shown in the top panels in Figure~\ref{fig:BayDetails}, which demonstrates our procedure for two bays. The smoothed image is then passed through a Sobel filter to extract gradients. The Sobel-processed image is the basis for a Canny edge identifier and contouring tool, producing a list of identified contours (Middle panels in Figure~\ref{fig:BayDetails}). From the identified edges, we selected elongated contours that best fit the bay - both in the smoothed version and the Sobel-processed image. These are then fitted with a parabola equation of the following form:
\begin{equation}
    \left[\frac{k(y-y0)+(x-x0)}{\sqrt{1+k^2}}\right]^2 = a\frac{y-y0-k(x-x0)}{\sqrt{1+k^2}} ,
\end{equation}
where $(x,y)$ are the image coordinates in pixel units, $k$ is related to the rotation, $(x0,y0)$ translation and $a$ the width of the fitted parabola. These parameters are then used to define the direction and focal point of the parabola: $(x_p,y_p)$ (red asterisk in the bottom panels of Figure~\ref{fig:BayDetails}).
We use the differential-evolution optimization method as implemented in the SciPy \citep{2020SciPy-NMeth} package to fit the parabola parameters.
For bays where the Sobel-processed image did not capture the bay outline satisfactorily, or the resulting contours were too short or many, manually selected points were added/used to supplement/enable parabolae fitting (Right column panels in Figure~\ref{fig:BayDetails}).

\begin{figure*}
\begin{center}
\includegraphics[trim=0cm 0cm 0cm 0cm,width=\textwidth]{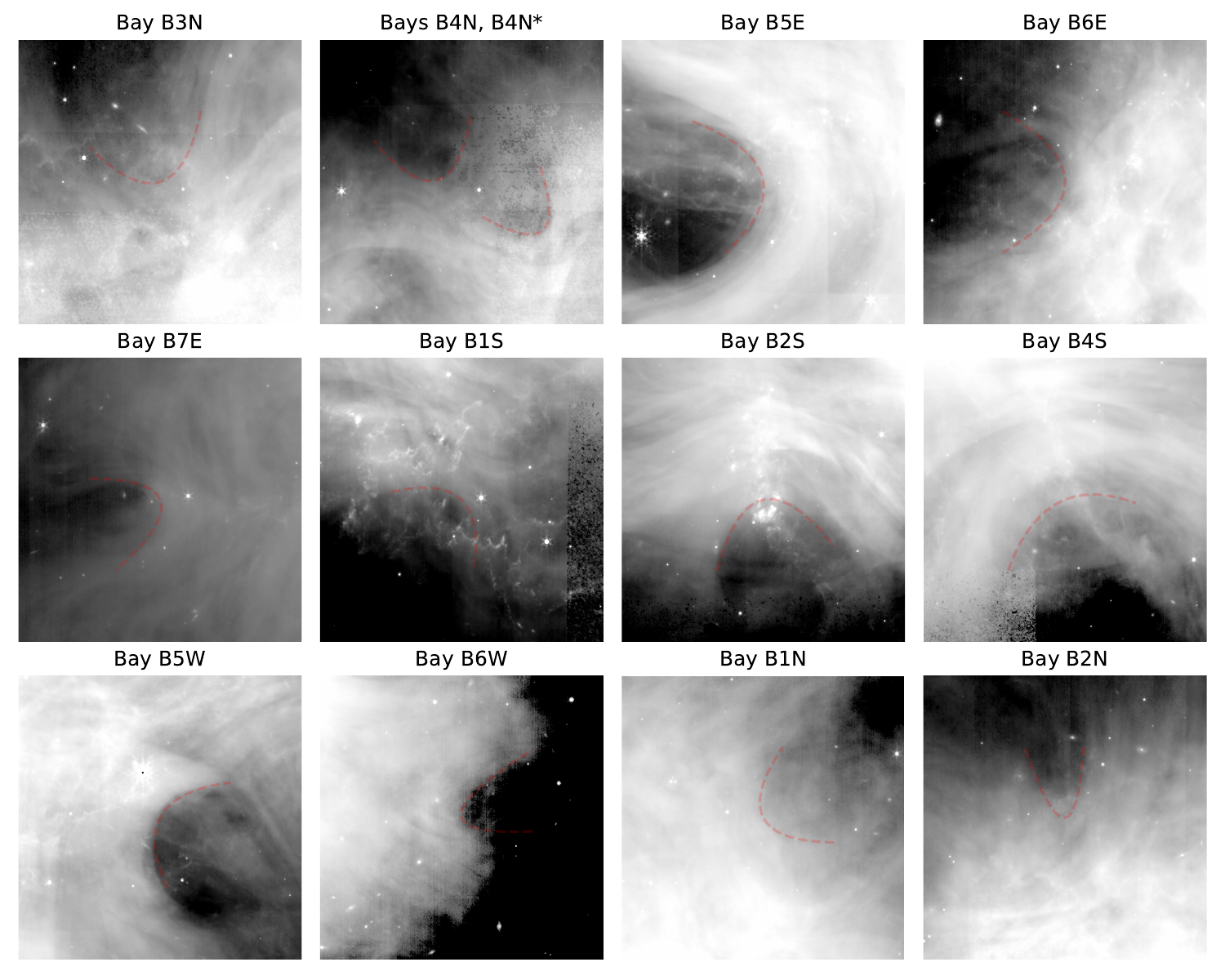} 
\caption{Zoom-in on the twelve additional bays, not depicted in Figure~\ref{fig:BayDetails}, with bay's name on top of each panel. In the panel denoted by "Bays B4N, B4N*" are two identified bays. For each bay region (JWST image), we draw the fitted parabola to the relevant bay with a red dashed line.}
\label{fig:BaysCloseups}
\end{center}
\end{figure*}

In Figure~\ref{fig:BaysCloseups}, we display a close-up of the remaining 12 bays and draw the fitted parabolae following the same procedure as described above and visually in Figure~\ref{fig:BayDetails}. In one of the panels in Figure~\ref{fig:BaysCloseups}, we identify and fit two bays. While the original marking from \cite{Temimetal2024} is consistent with both, Bay B4N (upper left) is more in line with the description in \cite{Temimetal2024}, which emphasized the possible link to dust filaments (Figure~\ref{fig:Temim2024}).

\begin{figure*}
\begin{center}
\includegraphics[trim=0cm 0cm 0cm 0cm,width=\textwidth]{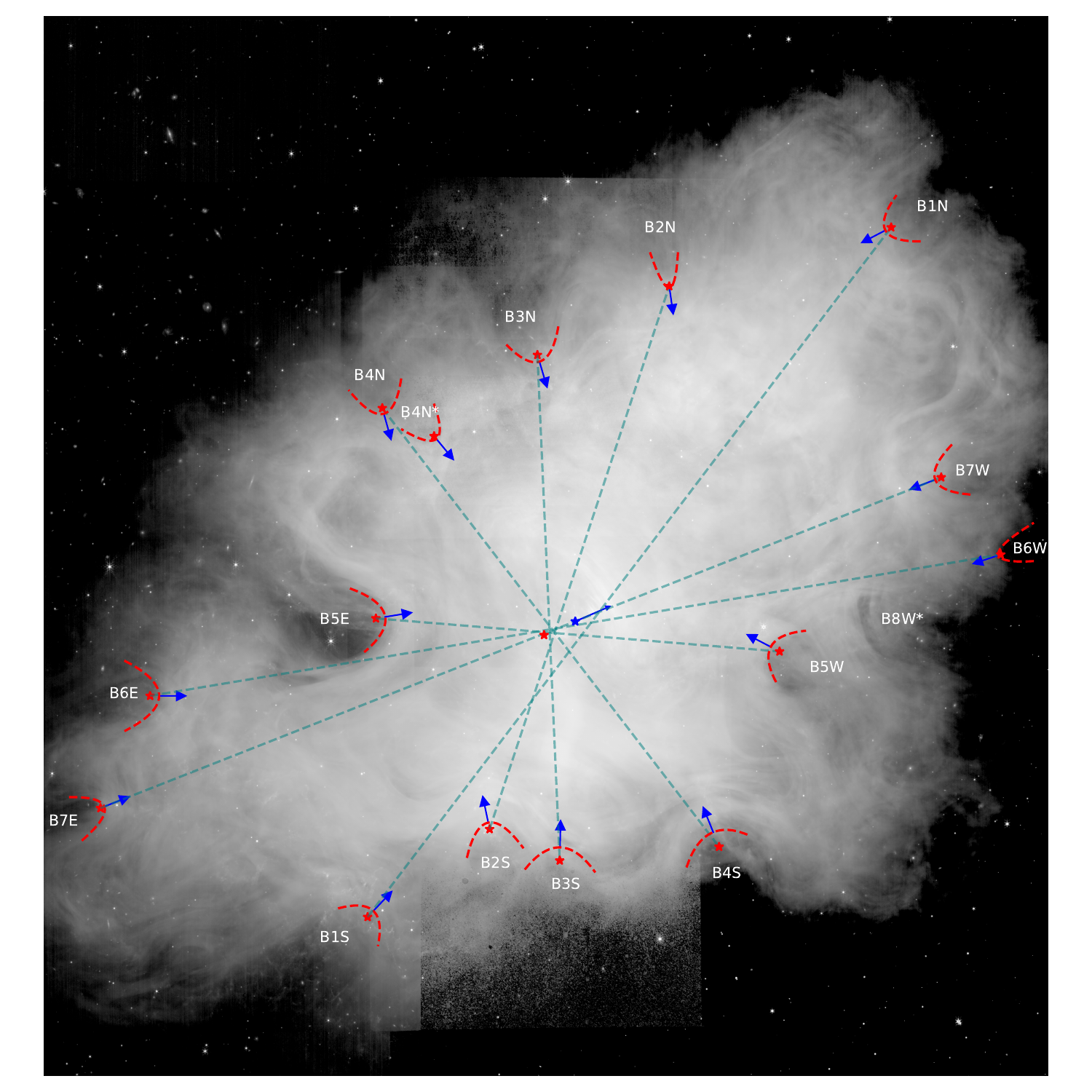} 
\caption{The JWST image with the identified bays, the parabolae we fit to them (Figures \ref{fig:BayDetails} and \ref{fig:BaysCloseups}), and the symmetry axes (dashed-teal) connecting their focal points. As in Figure~\ref{fig:Temim2024}, we denote with a blue asterisk the current location of the NS, with a red asterisk the deduced explosion site, and with a blue arrow the NS direction.}
\label{fig:Bays}
\end{center}
\end{figure*}

In connecting two opposite bays, we take the focal points of the parabolae as the edge of the symmetry axis. In Figure~\ref{fig:Bays}, we display the 14 identified bays, their focal points, their directions (symmetry line of each parabola; see Figure \ref{fig:BayDetails}), and the symmetry axis of each pair. We recall that another symmetry axis connects two filaments without clear bays, the dashed-red line in Figure~\ref{fig:Temim2024}.
Overall, we identify eight symmetry axes that compose the point-symmetrical morphology of the Crab Nebula.  

Six symmetry lines of bays (B2 through B7; Figures~\ref{fig:Temim2024} and \ref{fig:Bays}) and the symmetry line of the filaments (fE-fW; Figure~\ref{fig:Temim2024}) cross very close to the NS location at explosion (red asterisk). The symmetry line of pair B1 misses that location and the center of the line is largely displaced to the northwest. We propose three alternative explanations. 
(1) Our identification of bays and their pairing is incomplete; the counterpart to B1N might be beyond the extent of the image we used, and the counterpart to B1S might be obscured within the (optical) synchrotron nebula. (2) The pair B1 was shaped by the last pair of jets the NS launched, and this launching occurred after the NS acquired its full natal kick velocity. The NS kick velocity explains then the displacement of the center of the symmetry axis of pair B1 to the northwest, more or less the kick velocity direction. This possible explanation should be examined and confirmed by hydrodynamical simulations.
(3) Instabilities during the explosion process displaced the bays. We do not know how instabilities can cause this displacement, but we encourage future studies.

\section{Discussion and Summary}
\label{sec:Summary}

Starting with the nine bays that \cite{Temimetal2024} identified in a JWST image of the Crab Nebula and adding five bays that we identified here (Figure~\ref{fig:Temim2024}), we revealed the point-symmetrical morphology of the Crab. 
There are seven symmetry axes of bays, three pairs formed by six bays that \cite{Temimetal2024} identified, and one of a pair of filaments. These eight symmetry axes compose the point-symmetrical morphology of the Crab, as we draw in the right panel of Figure~\ref{fig:Temim2024}. 
In Section \ref{sec:Bays} we describe the fitting scheme we used for the identified bays and the resulting parabolae curves which define them. In each of the 14 locations, 9 by \cite{Temimetal2024} and 5 by us, we identify and fit at least one bay according to the JWST image and/or the Sobel-processed image (Section~\ref{sec:Bays}). We use the focal points of the parabolae to connect opposing bays and demonstrate that six of the lines pass in close proximity to the projected location of the NS at explosion (Figure~\ref{fig:Bays}), and offer explanations for the displaced seventh line at the end of Section~\ref{sec:Bays}.

\cite{Michaeletal1991} and \cite{Fesen_etal_1992} suggested that the largest pair of bays is the flow of the ejecta around a torus; the two large bays (B5E and B5W) are the cross sections of the torus on the plane of the sky. \cite{Fesen_etal_1992} estimated the mass of the torus of their model to be $>0.1 M_\odot$ of pre-explosion equatorial mass loss. However, with the identification of at least 7 pairs of bays, this explanation becomes problematic to the other pairs because we do not expect to have 7 torii in different planes. Another challenge to the pre-explosion torii model is explaining why all torii are on the outskirts of the ejecta and why some opposite bays in some pairs are at significantly different distances from the explosion point. 

We attribute the point-symmetrical structure to shaping by jittering jets, i.e., the explosion was by the JJEM.
Many jets that explode the star are choked in the core, leaving no observable morphological imprints different from instabilities. Some pairs of jets might leave point-symmetrical morphological signatures. Such pairs of jets can form pairs of ears and clumps as in, e.g., the Vela CCSNR (e.g., \citealt{SokerShishkin2024Vela}) and Cassiopeia A \citealt{BearSoker2025Cas}). The key difference of the Crab from the other CCSNRs with identified point-symmetrical morphologies is the non-negligible energy that the pulsar deposited to the ejecta, $E_{\rm P}/E_{\rm exp}  \simeq 0.3-0.7$ (see Section \ref{sec:Introduction}). After the Crab's progenitor explosion, the pulsar powered a PWN that accelerated the ejecta (e.g., \citealt{BlondingChevalier2017, MeyerDMAetal2024}). The dense clumps' acceleration was lower, but not much lower, explaining why the bays are close to the outer boundary of the ejecta. The low-density ejecta flew around the clumps, forming the bays at the location of the clumps. The clumps, hence the bays, have maintained their point-symmetrical morphology.  This scenario is the subject of future hydrodynamical simulations. 

Identifying a point-symmetrical morphology consistent with the JJEM in the Crab Nebula shows that jittering jets explode even very low-energy CCSNe. This firmly adds to the establishment of the JJEM as the primary explosion mechanism of CCSNe

\section*{Acknowledgements}
A grant from the Pazy Foundation supported this research. 

This work is based on observations made with the NASA/ESA/CSA James Webb Space Telescope. The data were obtained from the Mikulski Archive for Space Telescopes at the Space Telescope Science Institute, which is operated by the Association of Universities for Research in Astronomy, Inc., under NASA contract NAS 5-03127 for JWST.


\begin{thebibliography}{}
    \expandafter\ifx\csname natexlab\endcsname\relax\def\natexlab#1{#1}\fi
    \providecommand{\url}[1]{\href{#1}{#1}}
    \providecommand{\dodoi}[1]{doi:~\href{http://doi.org/#1}{\nolinkurl{#1}}}
    \providecommand{\doeprint}[1]{\href{http://ascl.net/#1}{\nolinkurl{http://ascl.net/#1}}}
    \providecommand{\doarXiv}[1]{\href{https://arxiv.org/abs/#1}{\nolinkurl{https://arxiv.org/abs/#1}}}
    
    \bibitem[{{Andresen} {et~al.}(2024){Andresen}, {O'Connor}, {Andersen}, \& {Couch}}]{Andresenetal2024}
    {Andresen}, H., {O'Connor}, E.~P., {Andersen}, O.~E., \& {Couch}, S.~M. 2024, \aap, 687, A55, \dodoi{10.1051/0004-6361/202449776}
    
    \bibitem[{{Bear} {et~al.}(2024){Bear}, {Shishkin}, \& {Soker}}]{Bearetal2025}
    {Bear}, E., {Shishkin}, D., \& {Soker}, N. 2024, arXiv e-prints, arXiv:2409.11453, \dodoi{10.48550/arXiv.2409.11453}
    
    \bibitem[{{Bear} \& {Soker}(2025)}]{BearSoker2025Cas}
    {Bear}, E., \& {Soker}, N. 2025, \na, 114, 102307, \dodoi{10.1016/j.newast.2024.102307}
    
    \bibitem[{{Bietenholz} \& {Nugent}(2015)}]{BietenholzNugent2015}
    {Bietenholz}, M.~F., \& {Nugent}, R.~L. 2015, \mnras, 454, 2416, \dodoi{10.1093/mnras/stv2112}
    
    \bibitem[{{Blondin} \& {Chevalier}(2017)}]{BlondingChevalier2017}
    {Blondin}, J.~M., \& {Chevalier}, R.~A. 2017, \apj, 845, 139, \dodoi{10.3847/1538-4357/aa8267}
    
    \bibitem[{Bradski(2000)}]{opencv_library}
    Bradski, G. 2000, Dr. Dobb's Journal of Software Tools
    
    \bibitem[{{Burrows} {et~al.}(2024){Burrows}, {Wang}, {Vartanyan}, \& {Coleman}}]{Burrowsetal2024kick}
    {Burrows}, A., {Wang}, T., {Vartanyan}, D., \& {Coleman}, M. S.~B. 2024, \apj, 963, 63, \dodoi{10.3847/1538-4357/ad2353}
    
    \bibitem[{{Clark} {et~al.}(1983){Clark}, {Murdin}, {Wood}, {Gilmozzi}, {Danziger}, \& {Furr}}]{Clarketal1983}
    {Clark}, D.~H., {Murdin}, P., {Wood}, R., {et~al.} 1983, \mnras, 204, 415, \dodoi{10.1093/mnras/204.2.415}
    
    \bibitem[{{Dirson} \& {Horns}(2023)}]{DirsonHorns2023}
    {Dirson}, L., \& {Horns}, D. 2023, \aap, 671, A67, \dodoi{10.1051/0004-6361/202243578}
    
    \bibitem[{{Dubner} {et~al.}(2017){Dubner}, {Castelletti}, {Kargaltsev}, {Pavlov}, {Bietenholz}, \& {Talavera}}]{Dubner_VLA_2017}
    {Dubner}, G., {Castelletti}, G., {Kargaltsev}, O., {et~al.} 2017, \apj, 840, 82, \dodoi{10.3847/1538-4357/aa6983}
    
    \bibitem[{{Fesen} {et~al.}(1992){Fesen}, {Martin}, \& {Shull}}]{Fesen_etal_1992}
    {Fesen}, R.~A., {Martin}, C.~L., \& {Shull}, J.~M. 1992, \apj, 399, 599, \dodoi{10.1086/171951}
    
    \bibitem[{{Gaia Collaboration} {et~al.}(2023){Gaia Collaboration}, {Vallenari}, {Brown}, {Prusti}, {de Bruijne}, {Arenou}, {Babusiaux}, {Biermann}, {Creevey}, {Ducourant}, {Evans}, {Eyer}, {Guerra}, {Hutton}, {Jordi}, {Klioner}, {Lammers}, {Lindegren}, {Luri}, {Mignard}, {Panem}, {Pourbaix}, {Randich}, {Sartoretti}, {Soubiran}, {Tanga}, {Walton}, {Bailer-Jones}, {Bastian}, {Drimmel}, {Jansen}, {Katz}, {Lattanzi}, {van Leeuwen}, {Bakker}, {Cacciari}, {Casta{\~n}eda}, {De Angeli}, {Fabricius}, {Fouesneau}, {Fr{\'e}mat}, {Galluccio}, {Guerrier}, {Heiter}, {Masana}, {Messineo}, {Mowlavi}, {Nicolas}, {Nienartowicz}, {Pailler}, {Panuzzo}, {Riclet}, {Roux}, {Seabroke}, {Sordo}, {Th{\'e}venin}, {Gracia-Abril}, {Portell}, {Teyssier}, {Altmann}, {Andrae}, {Audard}, {Bellas-Velidis}, {Benson}, {Berthier}, {Blomme}, {Burgess}, {Busonero}, {Busso}, {C{\'a}novas}, {Carry}, {Cellino}, {Cheek}, {Clementini}, {Damerdji}, {Davidson}, {de Teodoro}, {Nu{\~n}ez Campos}, {Delchambre}, {Dell'Oro}, {Esquej},
      {Fern{\'a}ndez-Hern{\'a}ndez}, {Fraile}, {Garabato}, {Garc{\'\i}a-Lario}, {Gosset}, {Haigron}, {Halbwachs}, {Hambly}, {Harrison}, {Hern{\'a}ndez}, {Hestroffer}, {Hodgkin}, {Holl}, {Jan{\ss}en}, {Jevardat de Fombelle}, {Jordan}, {Krone-Martins}, {Lanzafame}, {L{\"o}ffler}, {Marchal}, {Marrese}, {Moitinho}, {Muinonen}, {Osborne}, {Pancino}, {Pauwels}, {Recio-Blanco}, {Reyl{\'e}}, {Riello}, {Rimoldini}, {Roegiers}, {Rybizki}, {Sarro}, {Siopis}, {Smith}, {Sozzetti}, {Utrilla}, {van Leeuwen}, {Abbas}, {{\'A}brah{\'a}m}, {Abreu Aramburu}, {Aerts}, {Aguado}, {Ajaj}, {Aldea-Montero}, {Altavilla}, {{\'A}lvarez}, {Alves}, {Anders}, {Anderson}, {Anglada Varela}, {Antoja}, {Baines}, {Baker}, {Balaguer-N{\'u}{\~n}ez}, {Balbinot}, {Balog}, {Barache}, {Barbato}, {Barros}, {Barstow}, {Bartolom{\'e}}, {Bassilana}, {Bauchet}, {Becciani}, {Bellazzini}, {Berihuete}, {Bernet}, {Bertone}, {Bianchi}, {Binnenfeld}, {Blanco-Cuaresma}, {Blazere}, {Boch}, {Bombrun}, {Bossini}, {Bouquillon}, {Bragaglia}, {Bramante}, {Breedt},
      {Bressan}, {Brouillet}, {Brugaletta}, {Bucciarelli}, {Burlacu}, {Butkevich}, {Buzzi}, {Caffau}, {Cancelliere}, {Cantat-Gaudin}, {Carballo}, {Carlucci}, {Carnerero}, {Carrasco}, {Casamiquela}, {Castellani}, {Castro-Ginard}, {Chaoul}, {Charlot}, {Chemin}, {Chiaramida}, {Chiavassa}, {Chornay}, {Comoretto}, {Contursi}, {Cooper}, {Cornez}, {Cowell}, {Crifo}, {Cropper}, {Crosta}, {Crowley}, {Dafonte}, {Dapergolas}, {David}, {David}, {de Laverny}, {De Luise}, {De March}, {De Ridder}, {de Souza}, {de Torres}, {del Peloso}, {del Pozo}, {Delbo}, {Delgado}, {Delisle}, {Demouchy}, {Dharmawardena}, {Di Matteo}, {Diakite}, {Diener}, {Distefano}, {Dolding}, {Edvardsson}, {Enke}, {Fabre}, {Fabrizio}, {Faigler}, {Fedorets}, {Fernique}, {Fienga}, {Figueras}, {Fournier}, {Fouron}, {Fragkoudi}, {Gai}, {Garcia-Gutierrez}, {Garcia-Reinaldos}, {Garc{\'\i}a-Torres}, {Garofalo}, {Gavel}, {Gavras}, {Gerlach}, {Geyer}, {Giacobbe}, {Gilmore}, {Girona}, {Giuffrida}, {Gomel}, {Gomez}, {Gonz{\'a}lez-N{\'u}{\~n}ez},
      {Gonz{\'a}lez-Santamar{\'\i}a}, {Gonz{\'a}lez-Vidal}, {Granvik}, {Guillout}, {Guiraud}, {Guti{\'e}rrez-S{\'a}nchez}, {Guy}, {Hatzidimitriou}, {Hauser}, {Haywood}, {Helmer}, {Helmi}, {Sarmiento}, {Hidalgo}, {Hilger}, {H{\l}adczuk}, {Hobbs}, {Holland}, {Huckle}, {Jardine}, {Jasniewicz}, {Jean-Antoine Piccolo}, {Jim{\'e}nez-Arranz}, {Jorissen}, {Juaristi Campillo}, {Julbe}, {Karbevska}, {Kervella}, {Khanna}, {Kontizas}, {Kordopatis}, {Korn}, {K{\'o}sp{\'a}l}, {Kostrzewa-Rutkowska}, {Kruszy{\'n}ska}, {Kun}, {Laizeau}, {Lambert}, {Lanza}, {Lasne}, {Le Campion}, {Lebreton}, {Lebzelter}, {Leccia}, {Leclerc}, {Lecoeur-Taibi}, {Liao}, {Licata}, {Lindstr{\o}m}, {Lister}, {Livanou}, {Lobel}, {Lorca}, {Loup}, {Madrero Pardo}, {Magdaleno Romeo}, {Managau}, {Mann}, {Manteiga}, {Marchant}, {Marconi}, {Marcos}, {Marcos Santos}, {Mar{\'\i}n Pina}, {Marinoni}, {Marocco}, {Marshall}, {Martin Polo}, {Mart{\'\i}n-Fleitas}, {Marton}, {Mary}, {Masip}, {Massari}, {Mastrobuono-Battisti}, {Mazeh}, {McMillan}, {Messina}, {Michalik},
      {Millar}, {Mints}, {Molina}, {Molinaro}, {Moln{\'a}r}, {Monari}, {Mongui{\'o}}, {Montegriffo}, {Montero}, {Mor}, {Mora}, {Morbidelli}, {Morel}, {Morris}, {Muraveva}, {Murphy}, {Musella}, {Nagy}, {Noval}, {Oca{\~n}a}, {Ogden}, {Ordenovic}, {Osinde}, {Pagani}, {Pagano}, {Palaversa}, {Palicio}, {Pallas-Quintela}, {Panahi}, {Payne-Wardenaar}, {Pe{\~n}alosa Esteller}, {Penttil{\"a}}, {Pichon}, {Piersimoni}, {Pineau}, {Plachy}, {Plum}, {Poggio}, {Pr{\v{s}}a}, {Pulone}, {Racero}, {Ragaini}, {Rainer}, {Raiteri}, {Rambaux}, {Ramos}, {Ramos-Lerate}, {Re Fiorentin}, {Regibo}, {Richards}, {Rios Diaz}, {Ripepi}, {Riva}, {Rix}, {Rixon}, {Robichon}, {Robin}, {Robin}, {Roelens}, {Rogues}, {Rohrbasser}, {Romero-G{\'o}mez}, {Rowell}, {Royer}, {Ruz Mieres}, {Rybicki}, {Sadowski}, {S{\'a}ez N{\'u}{\~n}ez}, {Sagrist{\`a} Sell{\'e}s}, {Sahlmann}, {Salguero}, {Samaras}, {Sanchez Gimenez}, {Sanna}, {Santove{\~n}a}, {Sarasso}, {Schultheis}, {Sciacca}, {Segol}, {Segovia}, {S{\'e}gransan}, {Semeux}, {Shahaf}, {Siddiqui}, {Siebert},
      {Siltala}, {Silvelo}, {Slezak}, {Slezak}, {Smart}, {Snaith}, {Solano}, {Solitro}, {Souami}, {Souchay}, {Spagna}, {Spina}, {Spoto}, {Steele}, {Steidelm{\"u}ller}, {Stephenson}, {S{\"u}veges}, {Surdej}, {Szabados}, {Szegedi-Elek}, {Taris}, {Taylor}, {Teixeira}, {Tolomei}, {Tonello}, {Torra}, {Torra}, {Torralba Elipe}, {Trabucchi}, {Tsounis}, {Turon}, {Ulla}, {Unger}, {Vaillant}, {van Dillen}, {van Reeven}, {Vanel}, {Vecchiato}, {Viala}, {Vicente}, {Voutsinas}, {Weiler}, {Wevers}, {Wyrzykowski}, {Yoldas}, {Yvard}, {Zhao}, {Zorec}, {Zucker}, \& {Zwitter}}]{GAIA_DR3_2023}
    {Gaia Collaboration}, {Vallenari}, A., {Brown}, A.~G.~A., {et~al.} 2023, \aap, 674, A1, \dodoi{10.1051/0004-6361/202243940}
    
    \bibitem[{{Gardner} {et~al.}(2023){Gardner}, {Mather}, {Abbott}, {Abell}, {Abernathy}, {Abney}, {Abraham}, {Abraham}, {Abul-Huda}, {Acton}, {Adams}, {Adams}, {Adler}, {Adriaensen}, {Aguilar}, {Ahmed}, {Ahmed}, {Ahmed}, {Albat}, {Albert}, {Alberts}, {Aldridge}, {Allen}, {Allen}, {Altenburg}, {Altunc}, {Alvarez}, {{\'A}lvarez-M{\'a}rquez}, {Alves de Oliveira}, {Ambrose}, {Anandakrishnan}, {Andersen}, {Anderson}, {Anderson}, {Anderson}, {Anderson}, {Aprea}, {Archer}, {Arenberg}, {Argyriou}, {Arribas}, {Artigau}, {Arvai}, {Atcheson}, {Atkinson}, {Averbukh}, {Aymergen}, {Bacinski}, {Baggett}, {Bagnasco}, {Baker}, {Balzano}, {Banks}, {Baran}, {Barker}, {Barrett}, {Barringer}, {Barto}, {Bast}, {Baudoz}, {Baum}, {Beatty}, {Beaulieu}, {Bechtold}, {Beck}, {Beddard}, {Beichman}, {Bellagama}, {Bely}, {Berger}, {Bergeron}, {Bernier}, {Bertch}, {Beskow}, {Betz}, {Biagetti}, {Birkmann}, {Bjorklund}, {Blackwood}, {Blazek}, {Blossfeld}, {Bluth}, {Boccaletti}, {Boegner}, {Bohlin}, {Boia}, {B{\"o}ker}, {Bonaventura}, {Bond},
      {Bosley}, {Boucarut}, {Bouchet}, {Bouwman}, {Bower}, {Bowers}, {Bowers}, {Boyce}, {Boyer}, {Boyer}, {Boyer}, {Boyer}, {Bradley}, {Brady}, {Brandl}, {Brannen}, {Breda}, {Bremmer}, {Brennan}, {Bresnahan}, {Bright}, {Broiles}, {Bromenschenkel}, {Brooks}, {Brooks}, {Brown}, {Brown}, {Brown}, {Bruce}, {Bryson}, {Bujanda}, {Bullock}, {Bunker}, {Bureo}, {Burt}, {Bush}, {Bushouse}, {Bussman}, {Cabaud}, {Cale}, {Calhoon}, {Calvani}, {Canipe}, {Caputo}, {Cara}, {Carey}, {Case}, {Cesari}, {Cetorelli}, {Chance}, {Chandler}, {Chaney}, {Chapman}, {Charlot}, {Chayer}, {Cheezum}, {Chen}, {Chen}, {Cherinka}, {Chichester}, {Chilton}, {Chittiraibalan}, {Clampin}, {Clark}, {Clark}, {Clark}, {Claybrooks}, {Cleveland}, {Cohen}, {Cohen}, {Col{\'o}n}, {Coleman}, {Colina}, {Comber}, {Comeau}, {Comer}, {Conde Reis}, {Connolly}, {Conroy}, {Contos}, {Contreras}, {Cook}, {Cooper}, {Cooper}, {Correia}, {Correnti}, {Cossou}, {Costanza}, {Coulais}, {Cox}, {Coyle}, {Cracraft}, {Crew}, {Curtis}, {Cusveller}, {Da Costa Maciel}, {Dailey},
      {Daugeron}, {Davidson}, {Davies}, {Davis}, {Davis}, {Day}, {de Chambure}, {de Jong}, {De Marchi}, {Dean}, {Decker}, {Delisa}, {Dell}, {Dellagatta}, {Dembinska}, {Demosthenes}, {Dencheva}, {Deneu}, {DePriest}, {Deschenes}, {Dethienne}, {Detre}, {Diaz}, {Dicken}, {DiFelice}, {Dillman}, {Disharoon}, {Dixon}, {Doggett}, {Dominguez}, {Donaldson}, {Doria-Warner}, {Santos}, {Doty}, {Douglas}, {Doyon}, {Dressler}, {Driggers}, {Driggers}, {Dunn}, {DuPrie}, {Dupuis}, {Durning}, {Dutta}, {Earl}, {Eccleston}, {Ecobichon}, {Egami}, {Ehrenwinkler}, {Eisenhamer}, {Eisenhower}, {Eisenstein}, {El Hamel}, {Elie}, {Elliott}, {Elliott}, {Engesser}, {Espinoza}, {Etienne}, {Etxaluze}, {Evans}, {Fabreguettes}, {Falcolini}, {Falini}, {Fatig}, {Feeney}, {Feinberg}, {Fels}, {Ferdous}, {Ferguson}, {Ferrarese}, {Ferreira}, {Ferruit}, {Ferry}, {Filippazzo}, {Firre}, {Fix}, {Flagey}, {Flanagan}, {Fleming}, {Florian}, {Flynn}, {Foiadelli}, {Fontaine}, {Fontanella}, {Forshay}, {Fortner}, {Fox}, {Framarini}, {Francisco}, {Franck}, {Franx},
      {Franz}, {Friedman}, {Friend}, {Frost}, {Fu}, {Fullerton}, {Gaillard}, {Galkin}, {Gallagher}, {Galyer}, {Garc{\'\i}a Mar{\'\i}n}, {Gardner}, {Garland}, {Garrett}, {Gasman}, {G{\'a}sp{\'a}r}, {Gastaud}, {Gaudreau}, {Gauthier}, {Geers}, {Geithner}, {Gennaro}, {Gerber}, {Gereau}, {Giampaoli}, {Giardino}, {Gibbons}, {Gilbert}, {Gilman}, {Girard}, {Giuliano}, {Gkountis}, {Glasse}, {Glassmire}, {Glauser}, {Glazer}, {Goldberg}, {Golimowski}, {Gonzaga}, {Gordon}, {Gordon}, {Goudfrooij}, {Gough}, {Graham}, {Grau}, {Green}, {Greene}, {Greene}, {Greenfield}, {Greenhouse}, {Greve}, {Greville}, {Grimaldi}, {Groe}, {Groebner}, {Grumm}, {Grundy}, {G{\"u}del}, {Guillard}, {Guldalian}, {Gunn}, {Gurule}, {Gutman}, {Guy}, {Guyot}, {Hack}, {Haderlein}, {Hagan}, {Hagedorn}, {Hainline}, {Haley}, {Hami}, {Hamilton}, {Hammann}, {Hammel}, {Hanley}, {Hansen}, {Hardy}, {Harnisch}, {Harr}, {Harris}, {Hart}, {Hartig}, {Hasan}, {Hashim}, {Hashimoto}, {Haskins}, {Hawkins}, {Hayden}, {Hayden}, {Healy}, {Hecht}, {Heeg}, {Hejal}, {Helm},
      {Hengemihle}, {Henning}, {Henry}, {Henry}, {Henshaw}, {Hernandez}, {Herrington}, {Heske}, {Hesman}, {Hickey}, {Hilbert}, {Hines}, {Hinz}, {Hirsch}, {Hitcho}, {Hodapp}, {Hodge}, {Hoffman}, {Holfeltz}, {Holler}, {Hoppa}, {Horner}, {Howard}, {Howard}, {Huber}, {Hunkeler}, {Hunter}, {Hunter}, {Hurd}, {Hurst}, {Hutchings}, {Hylan}, {Ignat}, {Illingworth}, {Irish}, {Isaacs}, {Jackson}, {Jaffe}, {Jahic}, {Jahromi}, {Jakobsen}, {James}, {James}, {James}, {Jamieson}, {Jandra}, {Jayawardhana}, {Jedrzejewski}, {Jeffers}, {Jensen}, {Joanne}, {Johns}, {Johnson}, {Johnson}, {Johnson}, {Johnson}, {Johnson}, {Johnson}, {Johnstone}, {Jollet}, {Jones}, {Jones}, {Jones}, {Jones}, {Jones}, {Jordan}, {Jordan}, {Jue}, {Jurkowski}, {Justis}, {Justtanont}, {Kaleida}, {Kalirai}, {Kalmanson}, {Kaltenegger}, {Kammerer}, {Kan}, {Kanarek}, {Kao}, {Karakla}, {Karl}, {Kassin}, {Kauffman}, {Kavanagh}, {Kelley}, {Kelly}, {Kendrew}, {Kennedy}, {Kenny}, {Keski-Kuha}, {Keyes}, {Khan}, {Kidwell}, {Kimble}, {King}, {King}, {Kinzel}, {Kirk},
      {Kirkpatrick}, {Klaassen}, {Klingemann}, {Klintworth}, {Knapp}, {Knight}, {Knollenberg}, {Knutsen}, {Koehler}, {Koekemoer}, {Kofler}, {Kontson}, {Kovacs}, {Kozhurina-Platais}, {Krause}, {Kriss}, {Krist}, {Kristoffersen}, {Krogel}, {Krueger}, {Kulp}, {Kumari}, {Kwan}, {Kyprianou}, {Labador}, {Labiano}, {Lafreni{\`e}re}, {Lagage}, {Laidler}, {Laine}, {Laird}, {Lajoie}, {Lallo}, {Lam}, {LaMassa}, {Lambros}, {Lampenfield}, {Lander}, {Langston}, {Larson}, {Larson}, {LaVerghetta}, {Law}, {Lawrence}, {Lee}, {Lee}, {Lee}, {Leisenring}, {Leveille}, {Levenson}, {Levi}, {Levine}, {Lewis}, {Lewis}, {Lewis}, {Libralato}, {Lidon}, {Liebrecht}, {Lightsey}, {Lilly}, {Lim}, {Lim}, {Ling}, {Link}, {Link}, {Lipinski}, {Liu}, {Lo}, {Lobmeyer}, {Logue}, {Long}, {Long}, {Long}, {Long}, {L{\'o}pez-Caniego}, {Lotz}, {Love-Pruitt}, {Lubskiy}, {Luers}, {Luetgens}, {Luevano}, {Lui}, {Lund}, {Lundquist}, {Lunine}, {L{\"u}tzgendorf}, {Lynch}, {MacDonald}, {MacDonald}, {Macias}, {Macklis}, {Maghami}, {Maharaja}, {Maiolino},
      {Makrygiannis}, {Malla}, {Malumuth}, {Manjavacas}, {Marini}, {Marrione}, {Marston}, {Martel}, {Martin}, {Martin}, {Martinez}, {Maschmann}, {Masci}, {Masetti}, {Maszkiewicz}, {Matthews}, {Matuskey}, {McBrayer}, {McCarthy}, {McCaughrean}, {McClare}, {McClare}, {McCloskey}, {McClurg}, {McCoy}, {McElwain}, {McGregor}, {McGuffey}, {McKay}, {McKenzie}, {McLean}, {McMaster}, {McNeil}, {De Meester}, {Mehalick}, {Meixner}, {Mel{\'e}ndez}, {Menzel}, {Menzel}, {Merz}, {Mesterharm}, {Meyer}, {Meyett}, {Meza}, {Midwinter}, {Milam}, {Miller}, {Miller}, {Miskey}, {Misselt}, {Mitchell}, {Mohan}, {Montoya}, {Moran}, {Morishita}, {Moro-Mart{\'\i}n}, {Morrison}, {Morrison}, {Morse}, {Moschos}, {Moseley}, {Mosier}, {Mosner}, {Mountain}, {Muckenthaler}, {Mueller}, {Mueller}, {Muhiem}, {M{\"u}hlmann}, {Mullally}, {Mullen}, {Munger}, {Murphy}, {Murray}, {Muzerolle}, {Mycroft}, {Myers}, {Myers}, {Myers}, {Myers}, {Myrick}, {Nagle}, {Nayak}, {Naylor}, {Neff}, {Nelan}, {Nella}, {Nguyen}, {Nguyen}, {Nickson}, {Nidhiry}, {Niedner},
      {Nieto-Santisteban}, {Nikolov}, {Nishisaka}, {Noriega-Crespo}, {Nota}, {O'Mara}, {Oboryshko}, {O'Brien}, {Ochs}, {Offenberg}, {Ogle}, {Ohl}, {Olmsted}, {Osborne}, {O'Shaughnessy}, {{\"O}stlin}, {O'Sullivan}, {Otor}, {Ottens}, {Ouellette}, {Outlaw}, {Owens}, {Pacifici}, {Page}, {Paranilam}, {Park}, {Parrish}, {Paschal}, {Patapis}, {Patel}, {Patrick}, {Pattishall}, {Paul}, {Paul}, {Pauly}, {Pavlovsky}, {Pe{\~n}a-Guerrero}, {Pedder}, {Peek}, {Pelham}, {Penanen}, {Perriello}, {Perrin}, {Perrine}, {Perrygo}, {Peslier}, {Petach}, {Peterson}, {Pfarr}, {Pierson}, {Pietraszkiewicz}, {Pilchen}, {Pipher}, {Pirzkal}, {Pitman}, {Player}, {Plesha}, {Plitzke}, {Pohner}, {Poletis}, {Pollizzi}, {Polster}, {Pontius}, {Pontoppidan}, {Porges}, {Potter}, {Prescott}, {Proffitt}, {Pueyo}, {Quispe Neira}, {Radich}, {Rager}, {Rameau}, {Ramey}, {Ramos Alarcon}, {Rampini}, {Rapp}, {Rashford}, {Rauscher}, {Ravindranath}, {Rawle}, {Rawlings}, {Ray}, {Regan}, {Rehm}, {Rehm}, {Reid}, {Reis}, {Renk}, {Reoch}, {Ressler}, {Rest},
      {Reynolds}, {Richon}, {Richon}, {Ridgaway}, {Riedel}, {Rieke}, {Rieke}, {Rifelli}, {Rigby}, {Riggs}, {Ringel}, {Ritchie}, {Rix}, {Robberto}, {Robinson}, {Robinson}, {Robinson}, {Rock}, {Rodriguez}, {Rodr{\'\i}guez del Pino}, {Roellig}, {Rohrbach}, {Roman}, {Romelfanger}, {Romo}, {Rosales}, {Rose}, {Roteliuk}, {Roth}, {Rothwell}, {Rouzaud}, {Rowe}, {Rowlands}, {Roy}, {Royer}, {Rui}, {Rumler}, {Rumpl}, {Russ}, {Ryan}, {Ryan}, {Saad}, {Sabata}, {Sabatino}, {Sabbi}, {Sabelhaus}, {Sabia}, {Sahu}, {Saif}, {Salvignol}, {Samara-Ratna}, {Samuelson}, {Sanders}, {Sappington}, {Sargent}, {Sauer}, {Savadkin}, {Sawicki}, {Schappell}, {Scheffer}, {Scheithauer}, {Scherer}, {Schiff}, {Schlawin}, {Schmeitzky}, {Schmitz}, {Schmude}, {Schneider}, {Schreiber}, {Schroeven-Deceuninck}, {Schultz}, {Schwab}, {Schwartz}, {Scoccimarro}, {Scott}, {Scott}, {Seaton}, {Seely}, {Seery}, {Seidleck}, {Sembach}, {Shanahan}, {Shaughnessy}, {Shaw}, {Shay}, {Sheehan}, {Sheth}, {Shih}, {Shivaei}, {Siegel}, {Sienkiewicz}, {Simmons}, {Simon},
      {Sirianni}, {Sivaramakrishnan}, {Slade}, {Sloan}, {Slocum}, {Slowinski}, {Smith}, {Smith}, {Smith}, {Smith}, {Smith}, {Smith}, {Smolik}, {Soderblom}, {Sohn}, {Sokol}, {Sonneborn}, {Sontag}, {Sooy}, {Soummer}, {Southwood}, {Spain}, {Sparmo}, {Speer}, {Spencer}, {Sprofera}, {Stallcup}, {Stanley}, {Stansberry}, {Stark}, {Starr}, {Stassi}, {Steck}, {Steeley}, {Stephens}, {Stephenson}, {Stewart}, {Stiavelli}, {}, {Strada}, {Straughn}, {Streetman}, {Strickland}, {Strobele}, {Stuhlinger}, {Stys}, {Such}, {Sukhatme}, {Sullivan}, {Sullivan}, {Sumner}, {Sun}, {Sunnquist}, {Swade}, {Swam}, {Swenton}, {Swoish}, {Tam Litten}, {Tamas}, {Tao}, {Taylor}, {Taylor}, {te Plate}, {Van Tea}, {Teague}, {Telfer}, {Temim}, {Texter}, {Thatte}, {Thompson}, {Thompson}, {Thomson}, {Thronson}, {Tierney}, {Tikkanen}, {Tinnin}, {Tippet}, {Todd}, {Tran}, {Trauger}, {Trejo}, {Vinh Truong}, {Tsukamoto}, {Tufail}, {Tumlinson}, {Tustain}, {Tyra}, {Ubeda}, {Underwood}, {Uzzo}, {Vaclavik}, {Valenduc}, {Valenti}, {Van Campen}, {van de Wetering},
      {Van Der Marel}, {van Haarlem}, {Vandenbussche}, {van Dishoeck}, {Vanterpool}, {Vernoy}, {Vila Costas}, {Volk}, {Voorzaat}, {Voyton}, {Vydra}, {Waddy}, {Waelkens}, {Wahlgren}, {Walker}, {Wander}, {Warfield}, {Warner}, {Wasiak}, {Wasiak}, {Wehner}, {Weiler}, {Weilert}, {Weiss}, {Wells}, {Welty}, {Wheate}, {Wheeler}, {White}, {Whitehouse}, {Whiteleather}, {Whitman}, {Williams}, {Willmer}, {Willott}, {Willoughby}, {Wilson}, {Wilson}, {Wilson}, {Windhorst}, {Wislowski}, {Wolfe}, {Wolfe}, {Wolff}, {Wondel}, {Woo}, {Woods}, {Worden}, {Workman}, {Wright}, {Wu}, {Wu}, {Wun}, {Wymer}, {Yadetie}, {Yan}, {Yang}, {Yates}, {Yeager}, {Yerger}, {Young}, {Young}, {Yu}, {Yu}, {Zak}, {Zeidler}, {Zepp}, {Zhou}, {Zincke}, {Zonak}, \& {Zondag}}]{Gardner_JWST}
    {Gardner}, J.~P., {Mather}, J.~C., {Abbott}, R., {et~al.} 2023, \pasp, 135, 068001, \dodoi{10.1088/1538-3873/acd1b5}
    
    \bibitem[{{Janka} \& {Kresse}(2024)}]{JankaKresse2024}
    {Janka}, H.~T., \& {Kresse}, D. 2024, arXiv e-prints, arXiv:2401.13817, \dodoi{10.48550/arXiv.2401.13817}
    
    \bibitem[{{Kaplan} {et~al.}(2008){Kaplan}, {Chatterjee}, {Gaensler}, \& {Anderson}}]{Kaplan_etal_2008}
    {Kaplan}, D.~L., {Chatterjee}, S., {Gaensler}, B.~M., \& {Anderson}, J. 2008, \apj, 677, 1201, \dodoi{10.1086/529026}
    
    \bibitem[{{Lin} {et~al.}(2023){Lin}, {van Kerkwijk}, {Kirsten}, {Pen}, \& {Deller}}]{Lin_VLBI_2023}
    {Lin}, R., {van Kerkwijk}, M.~H., {Kirsten}, F., {Pen}, U.-L., \& {Deller}, A.~T. 2023, \apj, 952, 161, \dodoi{10.3847/1538-4357/acdc98}
    
    \bibitem[{{Loll} {et~al.}(2013){Loll}, {Desch}, {Scowen}, \& {Foy}}]{Loll_HST_2013}
    {Loll}, A.~M., {Desch}, S.~J., {Scowen}, P.~A., \& {Foy}, J.~P. 2013, \apj, 765, 152, \dodoi{10.1088/0004-637X/765/2/152}
    
    \bibitem[{{Martin} {et~al.}(2021){Martin}, {Milisavljevic}, \& {Drissen}}]{Martin_3D_2021}
    {Martin}, T., {Milisavljevic}, D., \& {Drissen}, L. 2021, \mnras, 502, 1864, \dodoi{10.1093/mnras/staa4046}
    
    \bibitem[{{Meyer} {et~al.}(2024){Meyer}, {Meliani}, \& {Torres}}]{MeyerDMAetal2024}
    {Meyer}, D.~M.~A., {Meliani}, Z., \& {Torres}, D.~F. 2024, arXiv e-prints, arXiv:2409.15829, \dodoi{10.48550/arXiv.2409.15829}
    
    \bibitem[{{Michel} {et~al.}(1991){Michel}, {Scowen}, {Dufour}, \& {Hester}}]{Michaeletal1991}
    {Michel}, F.~C., {Scowen}, P.~A., {Dufour}, R.~J., \& {Hester}, J.~J. 1991, \apj, 368, 463, \dodoi{10.1086/169710}
    
    \bibitem[{{M{\"u}ller} {et~al.}(2024){M{\"u}ller}, {Heger}, \& {Powell}}]{Mulleretal2024}
    {M{\"u}ller}, B., {Heger}, A., \& {Powell}, J. 2024, arXiv e-prints, arXiv:2407.08407, \dodoi{10.48550/arXiv.2407.08407}
    
    \bibitem[{{Nakamura} {et~al.}(2024){Nakamura}, {Takiwaki}, {Matsumoto}, \& {Kotake}}]{Nakamuraetal2024}
    {Nakamura}, K., {Takiwaki}, T., {Matsumoto}, J., \& {Kotake}, K. 2024, arXiv e-prints, arXiv:2405.08367, \dodoi{10.48550/arXiv.2405.08367}
    
    \bibitem[{{Porth} {et~al.}(2014){Porth}, {Komissarov}, \& {Keppens}}]{Porthetal2014}
    {Porth}, O., {Komissarov}, S.~S., \& {Keppens}, R. 2014, \mnras, 438, 278, \dodoi{10.1093/mnras/stt2176}
    
    \bibitem[{{Rieke} {et~al.}(2023){Rieke}, {Kelly}, {Misselt}, {Stansberry}, {Boyer}, {Beatty}, {Egami}, {Florian}, {Greene}, {Hainline}, {Leisenring}, {Roellig}, {Schlawin}, {Sun}, {Tinnin}, {Williams}, {Willmer}, {Wilson}, {Clark}, {Rohrbach}, {Brooks}, {Canipe}, {Correnti}, {DiFelice}, {Gennaro}, {Girard}, {Hartig}, {Hilbert}, {Koekemoer}, {Nikolov}, {Pirzkal}, {Rest}, {Robberto}, {Sunnquist}, {Telfer}, {Wu}, {Ferry}, {Lewis}, {Baum}, {Beichman}, {Doyon}, {Dressler}, {Eisenstein}, {Ferrarese}, {Hodapp}, {Horner}, {Jaffe}, {Johnstone}, {Krist}, {Martin}, {McCarthy}, {Meyer}, {Rieke}, {Trauger}, \& {Young}}]{Reike_NIRCam}
    {Rieke}, M.~J., {Kelly}, D.~M., {Misselt}, K., {et~al.} 2023, \pasp, 135, 028001, \dodoi{10.1088/1538-3873/acac53}
    
    \bibitem[{{Seward} {et~al.}(2006){Seward}, {Tucker}, \& {Fesen}}]{Seward_Chandra_2006}
    {Seward}, F.~D., {Tucker}, W.~H., \& {Fesen}, R.~A. 2006, \apj, 652, 1277, \dodoi{10.1086/508532}
    
    \bibitem[{{Soker}(2024{\natexlab{a}})}]{Soker2025W44}
    {Soker}, N. 2024{\natexlab{a}}, arXiv e-prints, arXiv:2411.04654, \dodoi{10.48550/arXiv.2411.04654}
    
    \bibitem[{{Soker}(2024{\natexlab{b}})}]{Soker2024CF}
    ---. 2024{\natexlab{b}}, The Open Journal of Astrophysics, 7, 49, \dodoi{10.33232/001c.120279}
    
    \bibitem[{{Soker}(2024{\natexlab{c}})}]{Soker2024Keyhole}
    ---. 2024{\natexlab{c}}, Research in Astronomy and Astrophysics, 24, 075006, \dodoi{10.1088/1674-4527/ad4fc2}
    
    \bibitem[{{Soker}(2024{\natexlab{d}})}]{Soker2025Learning}
    ---. 2024{\natexlab{d}}, arXiv e-prints, arXiv:2409.13657, \dodoi{10.48550/arXiv.2409.13657}
    
    \bibitem[{{Soker}(2024{\natexlab{e}})}]{Soker2024Rev}
    ---. 2024{\natexlab{e}}, The Open Journal of Astrophysics, 7, 31, \dodoi{10.33232/001c.117147}
    
    \bibitem[{{Soker} \& {Shishkin}(2024)}]{SokerShishkin2024Vela}
    {Soker}, N., \& {Shishkin}, D. 2024, arXiv e-prints, arXiv:2409.02626, \dodoi{10.48550/arXiv.2409.02626}
    
    \bibitem[{{Temim} {et~al.}(2006){Temim}, {Gehrz}, {Woodward}, {Roellig}, {Smith}, {Rudnick}, {Polomski}, {Davidson}, {Yuen}, \& {Onaka}}]{Temim_Spitzer_2006}
    {Temim}, T., {Gehrz}, R.~D., {Woodward}, C.~E., {et~al.} 2006, \aj, 132, 1610, \dodoi{10.1086/507076}
    
    \bibitem[{{Temim} {et~al.}(2024){Temim}, {Laming}, {Kavanagh}, {Smith}, {Slane}, {Blair}, {De Looze}, {Bucciantini}, {Jerkstrand}, {Gountanis}, {Sankrit}, {Milisavljevic}, {Rest}, {Lyutikov}, {DePasquale}, {Martin}, {Drissen}, {Raymond}, {Fox}, {Modjaz}, {Spitkovsky}, \& {Strolger}}]{Temimetal2024}
    {Temim}, T., {Laming}, J.~M., {Kavanagh}, P.~J., {et~al.} 2024, \apjl, 968, L18, \dodoi{10.3847/2041-8213/ad50d1}
    
    \bibitem[{Virtanen {et~al.}(2020)Virtanen, Gommers, Oliphant, Haberland, Reddy, Cournapeau, Burovski, Peterson, Weckesser, Bright, {van der Walt}, Brett, Wilson, Millman, Mayorov, Nelson, Jones, Kern, Larson, Carey, Polat, Feng, Moore, {VanderPlas}, Laxalde, Perktold, Cimrman, Henriksen, Quintero, Harris, Archibald, Ribeiro, Pedregosa, {van Mulbregt}, \& {SciPy 1.0 Contributors}}]{2020SciPy-NMeth}
    Virtanen, P., Gommers, R., Oliphant, T.~E., {et~al.} 2020, Nature Methods, 17, 261, \dodoi{10.1038/s41592-019-0686-2}
    
    \bibitem[{{Wang} {et~al.}(2024){Wang}, {Shishkin}, \& {Soker}}]{WangShishkinSoker2024}
    {Wang}, N. Y.~N., {Shishkin}, D., \& {Soker}, N. 2024, \apj, 969, 163, \dodoi{10.3847/1538-4357/ad487f}
    
    \bibitem[{{Wright} {et~al.}(2023){Wright}, {Rieke}, {Glasse}, {Ressler}, {Garc{\'\i}a Mar{\'\i}n}, {Aguilar}, {Alberts}, {{\'A}lvarez-M{\'a}rquez}, {Argyriou}, {Banks}, {Baudoz}, {Boccaletti}, {Bouchet}, {Bouwman}, {Brandl}, {Breda}, {Bright}, {Cale}, {Colina}, {Cossou}, {Coulais}, {Cracraft}, {De Meester}, {Dicken}, {Engesser}, {Etxaluze}, {Fox}, {Friedman}, {Fu}, {Gasman}, {G{\'a}sp{\'a}r}, {Gastaud}, {Geers}, {Glauser}, {Gordon}, {Greene}, {Greve}, {Grundy}, {G{\"u}del}, {Guillard}, {Haderlein}, {Hashimoto}, {Henning}, {Hines}, {Holler}, {Detre}, {Jahromi}, {James}, {Jones}, {Justtanont}, {Kavanagh}, {Kendrew}, {Klaassen}, {Krause}, {Labiano}, {Lagage}, {Lambros}, {Larson}, {Law}, {Lee}, {Libralato}, {Lorenzo Alverez}, {Meixner}, {Morrison}, {Mueller}, {Murray}, {Mycroft}, {Myers}, {Nayak}, {Naylor}, {Nickson}, {Noriega-Crespo}, {{\"O}stlin}, {O'Sullivan}, {Ottens}, {Patapis}, {Penanen}, {Pietraszkiewicz}, {Ray}, {Regan}, {Roteliuk}, {Royer}, {Samara-Ratna}, {Samuelson}, {Sargent}, {Scheithauer},
      {Schneider}, {Schreiber}, {Shaughnessy}, {Sheehan}, {Shivaei}, {Sloan}, {Tamas}, {Teague}, {Temim}, {Tikkanen}, {Tustain}, {van Dishoeck}, {Vandenbussche}, {Weilert}, {Whitehouse}, \& {Wolff}}]{Wright_MIRI}
    {Wright}, G.~S., {Rieke}, G.~H., {Glasse}, A., {et~al.} 2023, \pasp, 135, 048003, \dodoi{10.1088/1538-3873/acbe66}
    
    \bibitem[{{Yang} \& {Chevalier}(2015)}]{YangChevalier2015}
    {Yang}, H., \& {Chevalier}, R.~A. 2015, \apj, 806, 153, \dodoi{10.1088/0004-637X/806/2/153}
    
    \end{thebibliography}
\end{document}